\documentclass[twocolumn,aps,showpacs,superscriptaddress,amsmath,amsfonts,amssymb,floatfix,pre]{revtex4}
\usepackage{graphicx}
\usepackage{dcolumn}
\usepackage{bm}

\begin{document}

\title{Universality in complex networks: random matrix analysis}

\author{Jayendra N. Bandyopadhyay}
\affiliation{Max-Planck Institute for the Physics of Complex Systems, 
N\"{o}thnitzerstr. 38, D-01187 Dresden, Germany}
\author{Sarika Jalan}
\affiliation{Max-Planck Institute for Mathematics in the 
Sciences, Inselstrasse 22, D-04103 Leipzig, Germany}

\begin{abstract}

We apply random matrix theory to complex networks. We show that
nearest neighbor spacing distribution of the eigenvalues of the
adjacency matrices of various model networks, namely scale-free, small-world
and random networks follow universal Gaussian orthogonal ensemble  
statistics of random matrix theory. Secondly we show an analogy between the 
onset of small-world behavior, quantified by the structural properties of 
networks, and the transition from Poisson to Gaussian orthogonal ensemble
statistics, quantified by Brody parameter characterizing a spectral property. 
We also present our analysis for a protein-protein interaction network in budding yeast. 

\end{abstract}


\maketitle

The network concept has been gaining recognition as a fundamental tool in 
understanding dynamical behavior and response of real systems 
coming from different fields such as biology (e.g. food-web, 
nervous system, cellular metabolism, protein-protein interaction network, 
gene regulatory networks), social systems (e.g. scientific collaboration, 
citation), linguistic networks, and technological systems (e.g. Internet, 
power-grid etc (for reviews, see, e.g., \cite{rev-network}).

Different models have been proposed to study and understand  
systems having underlying network structures. Watts and Strogatz proposed 
an algorithm to generate popularly known as `small-world network' 
\cite{SW}, which captures randomness (characterized by small 
diameter) and regularity (measured by clustering) of real-world networks. 
This model emphasizes on the importance of random connections
in networks. Barab\'asi and 
Albert proposed a model to capture degree distributions of  
real-world networks \cite{barabasi}. According to this model only few 
nodes are responsible to carry the whole network. Since then came spurt of
activities to the network studies and  various 
structural properties of these model networks and real world networks have been 
studied to a great extent \cite{rev-network,SW,barabasi,modular}.

Furthermore, there exists extensive literature demonstrating that  
the properties of networks are well characterized by the spectrum 
of associated adjacency matrices. The adjacency matrix ($A$) of a network is 
defined in the following way: $A_{ij} = 1$ if $i$ and $j$ nodes are connected 
and zero otherwise. For an undirected network it is 
symmetric and consequently has real eigenvalues. These eigenvalues give 
information about some basic topological properties of underlying 
networks \cite{book-Adj}. For example, spectral density of
adjacency matrix of a random network, whose elements are randomly $0$ or $1$,
also follows the semicircular law \cite{Vicsek}. Interestingly, this result
matches with a very celebrated result in RMT about the spectral density of a 
random matrix, whose 
elements are Gaussian distributed random numbers, following Wigner's 
semicircular law \cite{mehta}. 

With the increasing availability of large maps of real-world networks, the 
analysis of spectral densities of real-world 
networks and model networks having real-world properties have also begun 
\cite{Vicsek,Dorogovtsev,Aguiar}. These analyses show that the spectral 
densities of model networks and real-world networks are not semicircular, 
instead they have some specific features depending on the minute details 
of the networks . For example,  small-world model networks show 
very complex spectral densities with many sharp peaks, while spectral 
densities of scale-free model networks exhibit triangular distribution 
\cite{Vicsek,Aguiar}.

In this paper we study networks within the framework of random 
matrix theory (RMT). We show that there exists one to 
one correlation between the network diameter which is a structural property 
and the eigenvalues fluctuations of the adjacency matrix which is a spectral property. We 
present our RMT analysis for various model networks studied extensively in the recent 
network literature and also for a real-world network. We find that in 
spite of having differences (in terms of various local and global 
properties, which are being used to characterize networks) in these 
networks, fluctuations of the eigenvalues of adjacency matrices show 
universal distribution. So far we are aware of only one relevant paper 
where authors have studied eigenvalue fluctuations in a 
microarray data for discovering functional gene modules \cite{rmt-gene}. 

RMT was proposed by Wigner to explain statistical properties of 
nuclear spectra \cite{mehta}. Later this theory was successfully applied 
in the study of spectra of different complex systems including disordered systems, 
quantum chaotic systems, large complex atoms, etc 
\cite{weidenmueller}. More recently, RMT is applied successfully to analyze
time-series data of stock-market, atmosphere, human EEG, and many more 
\cite{time_series}. A popular practice in RMT is to study eigenvalue fluctuations
via nearest neighbor spacing distribution (NNSD). NNSD is the distribution of
spacings between consecutive eigenvalues. 
It follows two universal properties depending upon the underlying 
correlations among the eigenvalues. For correlated eigenvalues, 
NNSD follows Wigner-Dyson formula of Gaussian orthogonal ensemble (GOE) 
statistics of RMT, which is a property shown by real symmetric random matrices 
with elements being Gaussian distributed random numbers. On the other hand, for
uncorrelated eigenvalues, NNSD follows Poisson statistics of RMT, which is a
property shown by random matrices having nonzero elements only along its
diagonals.

In the present study, we find that the NNSD of random networks follow GOE. 
The spectral density of random networks and of Gaussian distributed 
random matrices are both semicircular, so it was expected that their spacing 
distributions would be identical. However, very interestingly, NNSD of 
scale-free networks and small-world networks also follow 
GOE statistics. In addition to these model networks, we also analyze a 
protein-protein interaction network in budding yeast. We find that this 
real-world network is scale-free and its spacing distribution also follows 
GOE.

\begin{figure}[t]
\centerline{
\includegraphics[height=6cm,width=7cm]{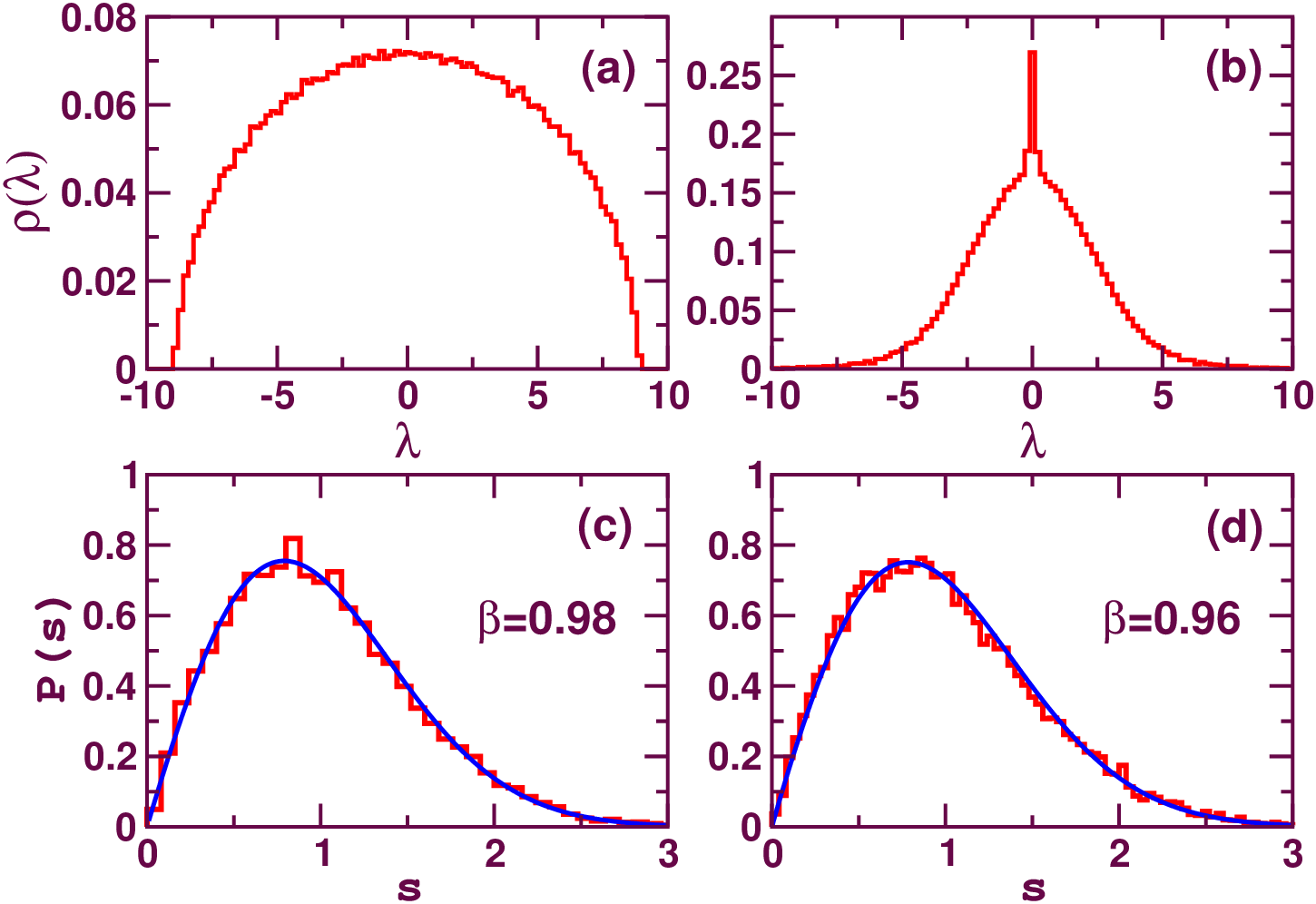}
}
\caption{(Color online) (a)-(b) Spectral density ($\rho(\lambda)$) of 
random (Erd\"{o}s-Renyi model) and scale-free network (following 
Ref.\cite{barabasi}), respectively. (c)-(d) Corresponding spacing 
distribution ($P(s)$). Both follow GOE statistics. The histograms are 
numerical results and the solid lines represent fitted Brody distribution. 
All networks have $N=2000$ nodes and an average degree $k = 20$ per node. 
Figures are plotted for average over 10 random realizations of the 
networks.}
\label{fig1}
\end{figure}

Secondly, we study the change of NNSD with the transition from regular to 
small-world network. Watts-Strogatz model of small-world network is 
constructed by rewiring the edges of a regular ring lattice with probability 
$p$. This rewiring procedure generates a network with some random 
connections, without altering the number of vertices or edges. For $p=0$, 
structure of the regular lattice or $k$-nearest neighbor coupled network 
remains same; on the other hand, for $p=1$, the regular lattice becomes 
a random network.  For intermediate values of $p$, the graph is a 
small-world network. We find that for the regular lattice ($p=0$), NNSD follows Poisson statistics,
for $p =1$ it follows GOE statistics and for $0< p < 1$ it shows intermediate 
statistics of Poisson and GOE. Moreover we show that the NNSD changes from 
Poisson to GOE with a very small increment in $p$, and most importantly, 
transition to GOE takes place exactly at the onset of small-world 
transition. We establish a relation between small-world transition 
and GOE transition by comparing the 
diameter and the clustering coefficients of network with the Brody parameter 
$\beta$.
This parameter comes from a semiempirical eigenvalues spacing distributions 
studied extensively in RMT to model Poisson to GOE transition.

Here we briefly describe some aspects of RMT which we use in our 
network analysis. We denote the eigenvalues of a network by 
$\lambda_i,\,\,i=1,\dots,N$, where $N$ is size of the network and 
$\lambda_1 < \lambda_2 < \lambda_3 < \dots < \lambda_N$. In 
order to get universal properties of the fluctuations of eigenvalues, 
it is customary in RMT to unfold the eigenvalues by a transformation 
$\overline{\lambda}_i = \overline{N} (\lambda_i)$, where $\overline{N}$ is 
averaged integrated eigenvalue density \cite{mehta}. Since we do not 
have any analytical form for $\overline{N}$, we numerically unfold the 
spectrum by polynomial curve fitting (for elaborate discussion on unfolding,
see Ref.\cite{mehta}). After unfolding, average 
spacings will be {\it unity}, independent of the system. Using the 
unfolded spectra, we calculate spacings as 
$s_i=\overline{\lambda}_{i+1}-\overline{\lambda}_i$. NNSD is 
defined as the probability distribution ($P(s)$) of these $s_i$'s. In case of 
Poisson statistics, $P(s)=\exp(-s)$; whereas for GOE, 
$P(s)=\frac{\pi}{2}s\exp \left(-\frac{\pi s^2}{4}\right)$. For  
intermediate cases, the spacing distribution is described by Brody 
distribution \cite{brody}:
\begin{subequations} 
\label{brody}
\begin{equation}
P_{\beta}(s) = A s^\beta\exp\left(-\alpha s^{\beta+1}\right),
\end{equation}
where
\begin{equation} 
A = (1+\beta)\alpha~\mbox{and}~ \alpha = 
\left[\Gamma \left( \frac{\beta+2}{\beta+1}\right)\right]^{\beta+1} 
\end{equation}
\end{subequations} 
This is a semiempirical formula characterized by parameter $\beta$. As 
$\beta$ goes from $0$ to $1$, the Brody distribution smoothly changes from 
Poisson to GOE. We fit spacing distributions of different networks by 
the Brody distribution $P_{\beta}(s)$. This fitting gives an estimation of 
$\beta$, and consequently identifies whether the spacing distribution of a 
given network is Poisson, GOE or intermediate of these {\it two}.

In Fig.~\ref{fig1}, we present the ensemble averaged spectral density 
($\rho(\lambda)$) and spacing distribution ($P(s)$) of random and 
the scale-free networks. Figs. \ref{fig1}(a) and \ref{fig1}(b) respectively 
show the well known semicircular and triangular distribution of  
spectral densities of random and scale-free networks. Using RMT techniques described
earlier we obtain spacing distributions for the unfolded eigenvalues. Figs. \ref{fig1}(c)
and \ref{fig1}(d) plot these distributions. Now using Eq.~(\ref{brody}), we estimate
Brody parameter as $\beta \simeq 1$, which clearly shows GOE statistics of  
spacing distributions
for both the networks.  Following RMT, these results 
imply that even though spectral densities of scale-free networks are 
different from random networks, correlations among the 
eigenvalues of scale-free networks are as strong as that of the random 
networks. 

\begin{figure}[t] 
\centerline{ 
\includegraphics[height=6cm,width=7cm]{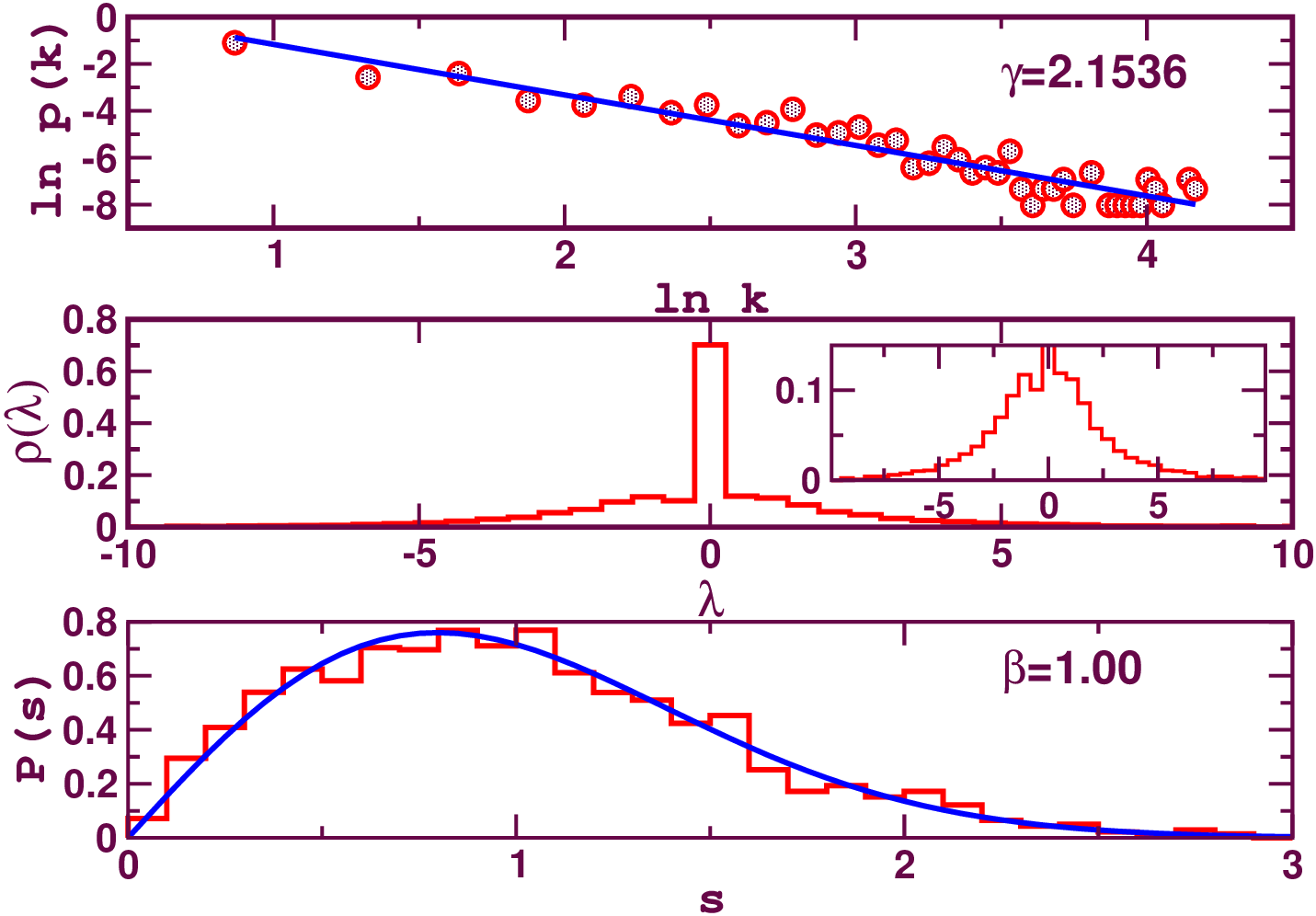} } \caption{(Color online) 
Figure shows different properties of a protein-protein interaction network 
in budding yeast. (a) Degree distribution : the scale-free nature of the 
network is clearly observed. (b) Spectral density : large value of 
$\rho(0)$ (Inset : besides large $\rho(0)$, overall spectral density 
follows well-known triangular distribution. (c) Spacing distribution : it 
follows GOE, estimated value of $\beta$ is $\sim 1$. The histogram 
represents numerical result and the solid line is fitted Brody 
distribution given by Eq.~(\ref{brody}).}
\label{yeast} 
\end{figure} 

\begin{figure}[b] \begin{center} 
\includegraphics[height=3cm,width=8cm]{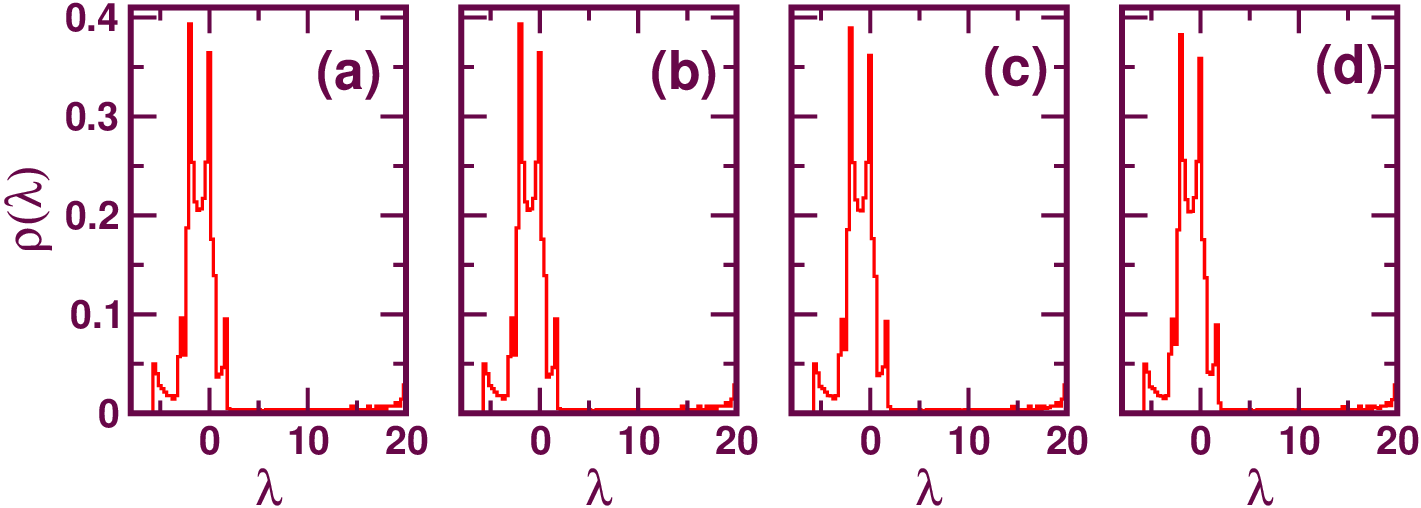}
\includegraphics[height=3cm,width=8cm]{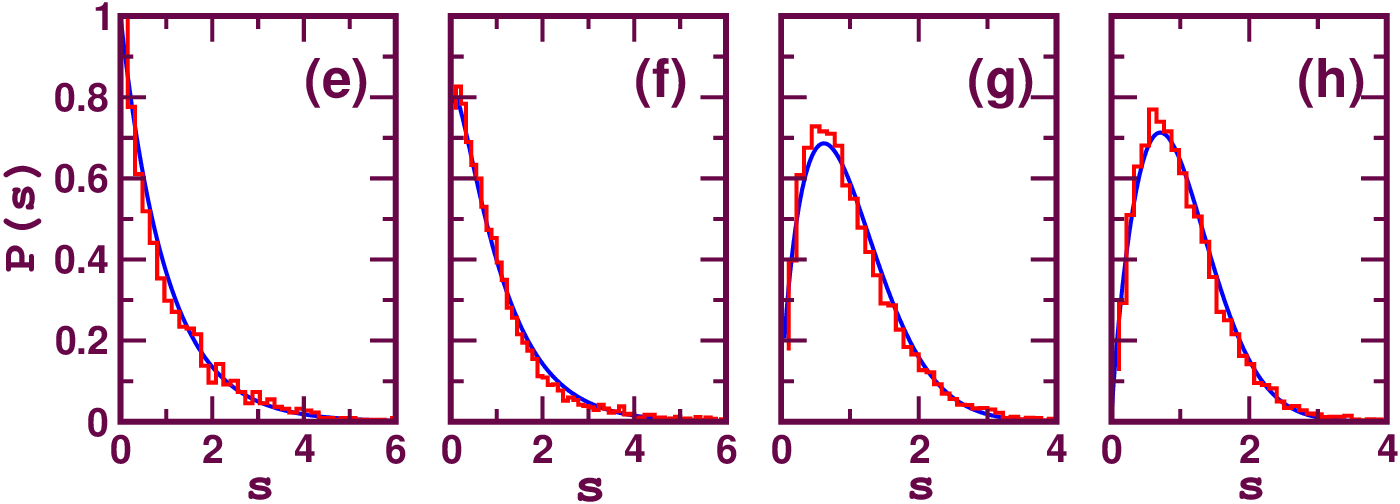}
\end{center}
\caption{(Color online) Figure shows the transition from ring regular 
lattice to the small-world network. (a)-(d) show the spectral densities 
and (e)-(h) show the corresponding spacing distributions for $p=0, 5 
\times 10^{-5}, 2 \times 10^{-4}, 5 \times 10^{-4}$, respectively. The 
histograms are numerical data and the solid lines are the corresponding 
fitted Brody distribution (Eq.~\ref{brody}). See text for the corresponding values of Brody 
parameters. All the networks have $N=2000$ nodes and $k=40$ average degree 
per node, and data are average over 10 random realization of the rewiring 
process.}
\label{sw}
\end{figure}

To show that our analysis exhibiting universality of GOE statistics for 
model random networks are generic, we studied some real-world networks 
also and here we present our results for a protein-protein interaction 
network in budding yeast \cite{budd_yeast}. Results are presented in 
Fig.~\ref{yeast}, top panel showing that the degree distribution $p(k)$ of 
the network follows power-law, i.e., $p(k) \propto k^{-\gamma}$, with 
$\gamma \simeq 2.1536$. Middle panel shows that the spectral 
density of this network is overall triangular (see also the inset of this 
panel for magnified figure) but with very large $\rho(0)$. Large value of 
$\rho(0)$ is one of the characteristics of many real-world networks 
\cite{Aguiar}. Due to the large $\rho(0)$, it is very difficult 
to numerically unfold the spectra. Therefore, in this case, we divide the 
spectra into two parts : one part contains only negative eigenvalues with 
values less than $-0.1$ and the other part contains positive eigenvalues 
with values greater than $0.1$. We assume these two sets of eigenvalues as 
an ensemble of two realizations, and calculate ensemble averaged spacing 
distribution. Bottom panel of Fig. \ref{yeast} is showing that the 
spacing distribution of this protein-protein interaction network  
follows GOE. We have also studied spectral rigidity of the eigenvalues
spectra which show that these networks follow RMT predictions for sufficiently 
large scales. Also this analysis seems to characterize the level of
randomness in network architecture \cite{pap3}. 

Now we discuss our results for the Watts-Strogatz model of small-world 
network. In Figs.~\ref{sw}(a) and \ref{sw}(e), we present respectively the 
spectral density and the spacing distribution of regular ring lattice with 
each node having $20$ edges. Spacing distributions are obtained again from the same technique.
Subfigure (a) shows that the spectral 
density of lattice is complicated without having any known analytical 
form; but its spacing distribution (subfigure (e)) clearly follows Poisson statistics 
($\beta \sim 0$). Then we randomize a fraction $p = 5 \times 10^{-5}$ of 
the edges of regular lattice. For this value of $p$, spectral 
density and spacing distribution are plotted respectively in Fig. 
\ref{sw}(b) and Fig. \ref{sw}(f). These figures reveal that, for this very 
small value of $p$, spectral density does not show any noticeable 
change as compared to the regular lattice, whereas 
spacing distribution shows different property ($\beta \sim 0.08$). 
As we further increase parameter $p$ from $5 \times 10^{-5}$ to $p = 2 
\times 10^{-4}$ and thereafter to $p = 5 \times 10^{-4}$, spectral 
densities show hardly any changes in its features (Figs. 
\ref{sw}(c)-\ref{sw}(d)), but very interestingly, according to Figs. 
\ref{sw}(g) and \ref{sw}(h), spacing distributions show significantly 
different properties as compared to the regular lattice. Now these are 
looking like 
intermediate of the Poisson and the GOE. By fitting spacing distribution 
corresponding to these two $p$ values with the Brody formula (Eq.~\ref{brody}), we estimate 
$\beta$ respectively as $0.63$ and $0.79$. These 
values indicate that we are already at the onset of 
Poisson$\rightarrow$GOE transition. 
Note that we take 
regular lattice with average degree $k \simeq 20$ for which
NNSD is showing Poisson statistics. For other values of $k$, where we 
may not have Poisson statistics, there also we get transition to GOE 
statistics. We choose $k \simeq 20$ just to make transition to GOE analogy clear.
Detailed analysis for other $k$ values would be presented elsewhere
\cite{pap-extend}.

\begin{figure}[t]
\centerline{
\includegraphics[height=5.5cm,width=6.5cm]{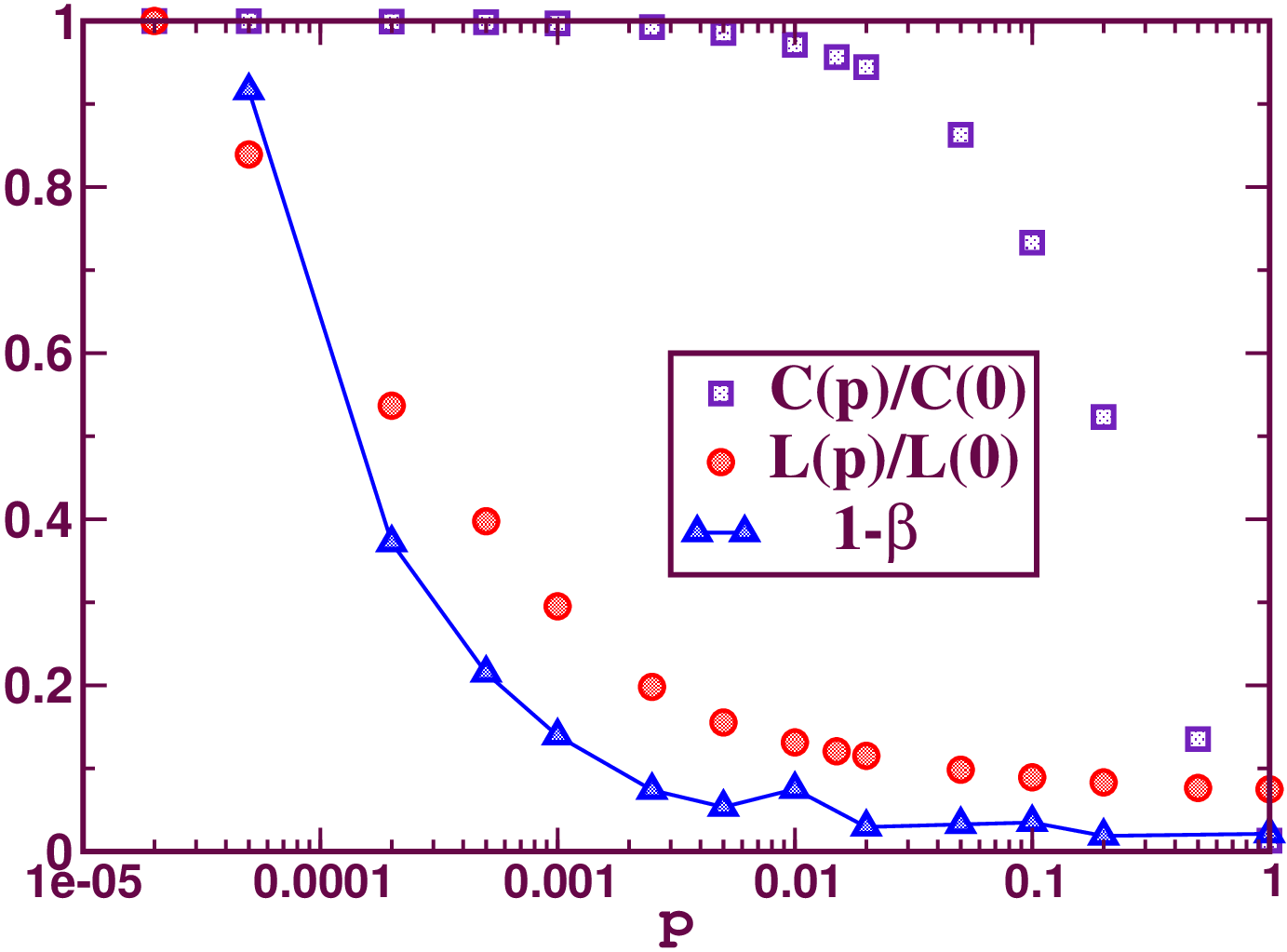}
}
\caption{(Color online) The shifted Brody parameter $1-\beta$ 
($\triangle$) is compared with the two well-known network parameters, 
normalized characteristic length $L(p)/L(0)$ ($\bigcirc$) and normalized 
clustering coefficients $C(p)/C(0)$ ($\square$). The data points 
corresponding to the curve for $\beta$ is joined by a solid line for 
better visibility. Network parameters are same as for the Fig. \ref{sw}. The 
data are average over 10 random realizations of rewiring process for each 
value of $p$.}
\label{beta_dia_clus}
\end{figure}

We present in Fig. \ref{beta_dia_clus} variation of  
$\beta$ as a function of $p$ over the whole range $0\le p \le 1$. 
Here we show correspondence between the Brody parameter and 
{\it two} important network parameters - the characteristic path length 
$L(p)$ and the clustering coefficient $C(p)$ - as a function of $p$. $L$ 
measures number of connections in the shortest path between two nodes, 
averaged over all pairs of the nodes. Clustering coefficient $C$ measures the 
cliquishness of a typical neighborhood, averaged over all nodes. In 
this figure we have normalized $L$ and $C$ by values $L(0)$ and $C(0)$ 
for the regular lattice. Due to this normalization, at $p=0$,  
normalized $L$ and $C$ both are {\it one} ; whereas for $p\rightarrow 1$, 
both network parameters will be closer to {\it zero}. However, $\beta$ 
behaves completely opposite way at the two extreme values of $p$. 
Therefore, in Fig. \ref{beta_dia_clus}, we compare $1-\beta$ with 
normalized the $L$ and $C$. This figure shows that the $\beta$ 
and the normalized characteristic length $L(p)/L(0)$ display similar 
trends and strong correspondence. The most important result of this study 
is that the Poisson$\rightarrow$GOE transition and the small-world 
transition take place at the same rewiring probability $p$. Note that  
all results presented here are for the adjacency matrices, however we have 
done similar analysis for Laplacian matrices also and for 
Fig. \ref{sw} and Fig. \ref{beta_dia_clus} qualitatively same results are 
obtained \cite{pap2}.

In summary, we study eigenvalues spacing distributions of various model 
networks and a real-world network. We show that though
the spectral densities of the random, the scale-free and the small-world
networks are different, their eigenvalues spacing distributions are same
and follow GOE statistics. We also show that spacing distribution for 
a protein-protein interaction network in budding yeast follows GOE statistics.
Following interpretation of RMT this universal
GOE statistics implies that the eigenvalues are strongly correlated 
among themselves because of some kind of randomness in the
corresponding matrix. In network concept this can be considered as 
sufficient amount of {\it randomness or disorder} in network 
connections. Furthermore, we study effect of randomness
in network architecture on the eigenvalues fluctuations,
and use Brody parameter to quantify this randomness.
We show that there exists one to one correlation 
between the network diameter, which is a structural property, and the 
Brody parameter characterizing a spectral property. We observe that {\it GOE 
transition} occurs at the onset of small-world transition. 
Again, this result implies that 
at the onset of small-world transition, there is some kind of {\it randomness}
spreading over the whole network leading to the strong correlations
among eigenvalues. The interesting point here is that
a very small amount of random connections is sufficient to give rise
these correlations.   

Now we point out some of the future prospects of our results. Universal GOE behavior of 
network spectra suggests that statistics of the bulk of eigenvalues of these
networks are consistent with those of a real symmetric random matrix with
entries being Gaussian distributed random numbers, and deviation from this
could be understood as system specific part.  
Random matrix analysis of eigenvectors had been performed for various different 
systems \cite{time_series}, such as stock-market, atmosphere, human EEG to extract system 
specific features by separating out 
universal properties from time-series of these systems. 
In same spirit, one can 
consider eigenvector analysis of adjacency matrices to understand
system specific features in different classes of networks \cite{pap-extend}. 
The system specific 
features could be important nodes, links or anything; but
the most important outcome of results presented in this paper is that
we can apply RMT, a very well developed branch of Physics, to study networks,
providing a completely new framework to the complex network research.

\begin{acknowledgments}
We thank Professor Steven Tomsovic for useful discussions and Dr. Luis G. Morelli (MPIPKS, Dresden) 
for useful suggestions.

\end{acknowledgments}


\begin{thebibliography}{99}

\bibitem{rev-network} R. Albert and A.-L. Barab\'asi, Rev. Mod. Phys.
{\bf 74}, 47 (2002) and references therein; 
S. Boccaletti {\it et al.}, Phys. Rep. {\bf 424}, 175 (2006).

\bibitem{SW} D. J. Watts and S. H. Strogatz, Nature {\bf 393}, 440 (1998).

\bibitem{barabasi} A.-L. Barab\'asi and R. Albert, Science {\bf 286}, 509 
(1999).

\bibitem{modular} E. Ravsaz {\it et al.}, Science {\bf 297}, 1551 (2002);
R. Guimer\'a and L. A. N. Amaral, Nature {\bf 433},
895 (2005).

\bibitem{book-Adj} D. M. Cvetkovi\'c, M. Doob and H. Sachs,
{\it Spectra of Graphs : theory and applications}, (Academic Press, 
3rd Revised edition, 1997).

\bibitem{Vicsek} I. J. Farkas {\it et al.}, Phys. Rev. E {\bf 64}, 026704, 
(2001).

\bibitem{mehta} M. L. Mehta, {\it Random Matrices}, 2nd ed. (Academic Press,
New York, 1991).

\bibitem{Dorogovtsev} S. N. Dorogovtsev {\it et al.}, Phys. Rev. E {\bf 68}, 
046109 (2003).

\bibitem{Aguiar} M. A. M. de Aguiar and Y. Bar-Yam, Phys. Rev. E
{\bf 71}, 016106 (2005).

\bibitem{rmt-gene} F. Luo {\it et al.}, Phys. Rev. E {\bf 73}, 031924 (2006).

\bibitem{weidenmueller} T. Guhr {\it et al.}, Phys. Rep. {\bf 299}, 189 (1998).

\bibitem{time_series} For examples, V. Pleron {\it et al.}, Phys. Rev. Lett.
{\bf 83}, 1471 (1999); M. S. Santhanam and P. K. Patra, Phys. Rev. E {\bf 64},
016102 (2001); P. Seba, Phys. Rev. Lett. {\bf 91}, 198104 (2003).

\bibitem{brody} T. A. Brody, Lett. Nuovo Cimento {\bf 7}, 482 (1973).

\bibitem{budd_yeast} http://vlado.fmf.uni-lj.si/pub/networks/data/bio
/Yeast/Yeast.htm.

\bibitem{pap3} S. Jalan and J. N. Bandyopadhyay, e-print : cond-mat/0701043.  
\bibitem{pap-extend} S. Jalan, J. N. Bandyopadhyay and M. S. Santhanam (under
preparation).

\bibitem{pap2} S. Jalan and J. N. Bandyopadhyay, e-print : cond-mat/0611735.  

\end{thebibliography}
\end{document}